\documentclass[11pt]{article}
\usepackage{amsmath,amssymb,amsthm,amsxtra,overpic}
\usepackage{hyperref}
\usepackage{url}
\begin{document}

\vspace{0.2cm}

\begin{center}
{\Large\bf Neutrino masses and flavor mixing in a generalized
inverse seesaw model with a universal two-zero texture}
\end{center}

\vspace{0.1cm}

\begin{center}
{\bf Ye-Ling Zhou}\footnote{E-mail: zhouyeling@ihep.ac.cn}
\\{Institute of High Energy Physics, Chinese Academy of
Sciences, Beijing 100049, China}
\end{center}

\vspace{0.8cm}

\begin{abstract}
A generalized inverse seesaw model, in which the $9\times9$ neutrino 
mass matrix has  vanishing (1,1) and (1,3) submatrices, is proposed.
This is similar to the universal two-zero texture which gives
vanishing (1,1) and (1,3) elements of the $3\times3$ mass matrices
in both the charged lepton and neutrino sectors. We consider the
$Z^{}_6\times Z^{}_6$ group to realize such texture zeros 
in the framework of the generalized inverse seesaw model. 
We also analyze the universal two-zero texture in the
general case and propose two ans$\ddot{\rm a}$tze to reduce the
number of free parameters. Taking account of the new result of
$\theta^{}_{13}$ from the Daya Bay experiment, we constrain the
parameter space of the universal two-zero texture in the general
case and in the two ans$\ddot{\rm a}$tze, respectively. We find that
one of the ans$\ddot{\rm a}$tze works well.
\end{abstract}

\begin{flushleft}
\hspace{0.8cm} PACS number(s): 14.60.Pq, 14.60.St, 98.80.Cq,
\end{flushleft}

\newpage
\section{Introduction}

The canonical seesaw mechanism \cite{seesaw} is successful in
generating small masses of left-handed neutrinos, but it has no
direct experimental testability and encounters a potential hierarchy
problem \cite{Xing09}. In the type-I seesaw model with heavy
right-handed neutrinos $N^{}_{\rm R}$, the left-handed neutrinos
$\nu^{}_{\rm L}$ can gain small masses $M^{}_{\nu} \approx M^{}_{\rm
D} M^{-1}_{\rm R} M^{T}_{\rm D}$ thanks to the huge right-handed
neutrino masses $M^{}_{\rm R}$. However, to obtain $M^{}_{\nu}\sim
{\rm \mathcal{O}(0.1) \;eV}$, one has to require $M^{}_{\rm R}\sim
\mathcal{O}(10^{14})\; {\rm GeV}$, if $M^{}_{\rm D}$ is assumed to
be at the electroweak scale ($\sim \mathcal{O}(10^2)\;{\rm GeV}$).
This makes the right-handed neutrinos far beyond the detectability
of any colliders. The hierarchy problem is that a very high seesaw
scale will lead to large corrections to the Higgs mass, which makes
the Higgs mass of the order of the electroweak scale unnatural. The
inverse seesaw model \cite{Mohapatra86} can solve these problems.
Moreover, it is possible to predict light sterile neutrinos
naturally \cite{Kim07} and provide rich phenomenology such as the
non-unitary effect and leptogenesis \cite{XZH09}.

The generalized inverse seesaw model (GISM) is an extension of the
canonical seesaw mechanism by introducing three right-handed
neutrinos $N^{}_{{\rm R}i}$ (for $i=1,2,3$), three additional
gauge-singlet neutrinos $S^{}_{{\rm R}i}$ and a scalar $\Phi$ into
the standard model (SM). The Lagrangian in the charged lepton and
neutrino sectors \cite{Ellis92} is written as
\begin{eqnarray}
-\mathcal{L}^{}_{l} &=& \overline{\ell^{}_{\rm L}} Y^{}_l H
E^{}_{\rm R} + \overline{\ell^{}_{\rm L}} Y^{}_{\rm D} \tilde{H}
N^{}_{\rm R} + \overline{N^{c}_{\rm R}} Y^{}_{\rm S} \Phi S^{}_{\rm
R} + \frac{1}{2} \overline{N^{c}_{\rm R}} M^{}_{\rm R} N^{}_{\rm R}
+ \frac{1}{2} \overline{S^{c}_{\rm R}} M^{}_{\mu} S^{}_{\rm R} +{\rm
h.c.} \;,
\end{eqnarray}
in which $H$, $\ell^{}_{\rm L}$ and $E^{}_{\rm R}$ stand for the
Higgs doublet, three lepton doublets and three charged-lepton
singlets, respectively in the SM and $\tilde{H}=i\sigma^{}_2
H^*_{}$. Here $Y^{}_l$, $Y^{}_{\rm D}$ and $Y^{}_{\rm S}$ are
$3\times3$ Yukawa coupling matrices, and $M^{}_{\rm R}$ and
$M^{}_{\mu}$ are $3\times3$ symmetric Majorana mass matrices. After
spontaneous symmetry breaking (SSB), the scalars acquire their
vacuum expectation values (VEVs), and we gain the $3\times3$ charged
lepton mass matrix $M^{}_{l}= Y^{}_l v(H)/\sqrt{2}$ and the
$9\times9$ neutrino mass matrix
\begin{eqnarray}
\mathcal{M} &=& \left( \begin{matrix} {\bf 0} & M^{}_{\rm D} & {\bf
0} \cr M^T_{\rm D} & M^{}_{\rm R} & M^{}_{\rm S} \cr {\bf 0} &
M^T_{\rm S} & M^{}_\mu
\end{matrix} \right)\;
\end{eqnarray}
in the flavor basis, in which $M^{}_{\rm D}= Y^{}_{\rm D}
v(H)/\sqrt{2}$ and $M^{}_{\rm S}= Y^{}_{\rm S} v(\Phi)/\sqrt{2}$.
Here $v(H)$ and $v(\Phi)$ are the VEVs of $H$ and $\Phi$,
respectively. The GISM degrades to the original inverse seesaw model
(OISM) when $M^{}_{\rm R}={\bf 0}$ is taken. It can also accommodate
a larger range of the sterile neutrino masses than the OISM
\cite{Kim07}.

If we regard each submatrix of $\mathcal{M}$ in Eq. (2) as a complex
number, we turn to a typical pattern of two-zero textures \cite{GX}.
Different from the models given in Ref. \cite{GX}, where $M^{}_l$ is
chosen to be diagonal and only $M^{}_\nu$ has the two-zero texture,
we propose the universal two-zero texture (UTZT) \cite{XZH03}, in
which both $M^{}_l$ and $M^{}_\nu$ have two-zero textures. As the
similar texture zeros of quark mass matrices can interpret the
smallness of flavor mixing angles in the quark sectors \cite{FX99},
we expect the UTZT will give us a better understanding of the lepton
flavor mixing. We write out the charged lepton and left-handed
neutrino mass matrices as
\begin{eqnarray}
M^{}_{l,\nu} &=& \left( \begin{matrix} 0 & A^{}_{l,\nu} & 0 \cr
A^{}_{l,\nu}   & C^{}_{l,\nu} & B^{}_{l,\nu} \cr 0 & B^{}_{l,\nu} &
D^{}_{l,\nu} \cr
\end{matrix} \right) \; .
\end{eqnarray}
Some work of this texture has been done in Ref. \cite{More,Hu11}.
Generally, texture zeros can be obtained from Abelian symmetries
\cite{Grimus04}. Later we will see that by means of these symmetries
the two-zero texture of $M^{}_{l}$ can be directly derived. For the light left-handed
neutrino matrix $M^{}_{\nu}$, if $M^{}_{\rm D}$, $M^{}_{\rm R}$,
$M^{}_{\rm S}$ and $M^{}_\mu$ all have the two-zero textures, which
will be a natural result from the symmetries, the seesaw mechanism
can guarantee that $M^{}_\nu$ achieves the two-zero texture
\cite{XZH03,Hu11}.

Recently, the Daya Bay collaboration reported a relatively large
$\theta^{}_{13}$ \cite{Dayabay} with its best-fit ($\pm1\sigma$
range) value $\theta^{}_{13}\simeq8.8^\circ_{}\pm0.8^\circ_{}$. It
is confirmed by the RENO experiment \cite{RENO}. The experimental
results of large $\theta^{}_{13}$ give us two motivations for the
UTZT. (1) Two phenomenological strategies towards understanding
lepton flavor mixing are outlined in Ref. \cite{Xing12}: the first
one is to start from a nearly constant flavor mixing pattern, and
the second one is to associate the mixing angles with the lepton
mass ratios. While it is a nontrivial job to generate a large
$\theta^{}_{13}$ from the first strategy according to flavor
symmetries, one may pay more attention to the second strategy. To
implement the second strategy, one generally requires some elements
of $M^{}_{l}$ and $M^{}_{\nu}$ to be zeros or sufficiently small
compared with their neighbors, and the two-zero texture is a typical
example of this kind. (2) As discussed in Ref. \cite{XZS11}, where
$M^{}_{l}$ is diagonal and $M^{}_\nu$ has a two-zero texture, it is
more likely to obtain a large $\theta^{}_{13}$ if $M^{}_\nu$ has
texture zeros as in Eq. (3) compared with the other texture zeros.
Taking advantage of this kind of texture zeros, we expand our
discussion to the scenario that both $M^{}_l$ and $M^{}_\nu$ have
such texture zeros. We expect that such texture can also gain a
large $\theta^{}_{13}$.

The rest of this paper is organized as follows. In section 2, we
propose a model to connect the GISM with the UTZT under the discrete
Abelian group $Z^{}_6\times Z^{}_6$. 
With this model, we can realize the two-zero textures of
$M^{}_{l}$, $M^{}_{\rm D}$, $M^{}_{\rm R}$, $M^{}_{\rm S}$ and
$M^{}_{\mu}$. 
However, the realization of the two-zero texture of $M^{}_\nu$ is a little non-trivial. 
Section 3 is devoted to see how the two-zero texture of $M^{}_\nu$ is realized.
In section 4, the UTZT is used to
explain the lepton flavor mixing, especially for large
$\theta^{}_{13}$. Both analytical and numerical results are
presented. The predictions for the effective masses in the tritium
beta decay and neutrinoless double beta ($0\nu2\beta$) decay are
also given in this section. Since the UTZT in the general case has
several adjustable parameters, it does not get stringent
experimental constraints. In section 5, we consider two
ans$\ddot{\rm a}$tze of the UTZT to constrain the parameter space.
Ansatz (A) is a natural approximation based on our model built in
section 2, and ansatz (B) is a special case which has been
considered in Ref. \cite{XZH03}. Section 6 is the conclusion of our
paper.

\section{A model connecting the UTZT with the GISM}

In this section, we illustrate a way to connect the GISM with the
UTZT. We rewrite the Lagrangian in the charged lepton and neutrino
sectors as
\begin{eqnarray}
-\mathcal{L}^{}_{l} &=& \overline{\ell^{}_{{\rm L}i}}
(Y^{a}_{l})^{}_{ij} H^{a}_{} E^{}_{{\rm R}j} +
\overline{\ell^{}_{{\rm L}i}} (Y^{a}_{\rm D})^{}_{ij}
\tilde{H}^{a}_{} N^{}_{{\rm R}j} + \overline{N^{c}_{{\rm R}i}}
(Y^{a}_{\rm S})^{}_{ij} \Phi^{a}_{} S^{}_{{\rm R}j} \nonumber\\
&& + \frac{1}{2} \overline{N^{c}_{{\rm R}i}} (Y^{a}_{\rm R})^{}_{ij}
\chi^{a}_{} N^{}_{{\rm R}j} + \frac{1}{2} \overline{S^{c}_{{\rm
R}i}} (Y^{a}_{\mu})^{}_{ij} \phi^{a}_{} S^{}_{{\rm R}j} +{\rm h.c.}
\;,
\end{eqnarray}
in which the repeated indices are summed. In our model, we introduce
three scalars into each term, so $a=1,2,3$. Comparing Eq. (4) with
Eq. (1), we can see that some replacements have been done.
$Y^{}_{l}H$, $Y^{}_{\rm D} \tilde{H}$, $Y^{}_{\rm S} \Phi$ are
replaced by $Y^{a}_{l} H^{a}_{}$, $Y^{a}_{\rm D} \tilde{H}^{a}_{}$,
$Y^{a}_{\rm S} \Phi^{a}_{}$ , respectively, and the scalars
$\chi^{a}_{}$, $\phi^{a}_{}$ are introduced to give the Majorana
masses of $N^{}_{\rm R}$, $S^{}_{\rm R}$, respectively. The purpose
to do these replacements has nothing to do with the GISM but to give
the two-zero textures of the mass matrices $M^{}_l$, $M^{}_{\rm D}$,
$M^{}_{\rm S}$, $M^{}_{\rm R}$ and $M^{}_{\mu}$.

A model for connecting the GISM with the UTZT can be built based on
a direct product of groups $G^{}_{1q^{}_1} \times G^{}_{2q^{}_2}
\equiv \mathcal{G}$:
\begin{itemize}
\item Each fermion or scalar transforms
under the group $G^{}_1$ with a charge $q^{}_1$. This rule aims to
realize the GISM. Since it is flavor-blind, different flavors in the
same multiplet (e.g., $N^{}_i$ and $N^{}_j$ with $i\neq j$) have the
same charges $q^{}_1$, and different scalars in the same Yukawa
coupling (e.g., $H^{a}_{}$ and $H^{b}_{}$ with $a\neq b$) have the
same charges $q^{}_1$, too.

\item Each fermion or scalar transforms under group
$G^{}_2$ with a charge $q^{}_2$. We choose $G^{}_2$ to be an Abelian
group $Z^{}_n$ to give the UTZT. In this case, different flavors in
the same multiplet should have different charges $q^{}_2$, and
different scalars in the same Yukawa coupling term should also have
different charges $q^{}_2$.
\end{itemize}
Generally speaking, there are many possibilities to choose $G^{}_1$
and $G^{}_2$, and it is essentially unnecessary to require that they
be equal to each other. Nevertheless, in view of the similar
structures of $\mathcal{M}$ and $M^{}_{l,\nu}$, we assume
$G^{}_1=G^{}_2=Z^{}_n$.

\begin{table}[t]
\caption{The charges of the fermions and scalars under
$Z^{}_{6q^{}_1}$.}
\begin{center}
\begin{tabular}{c|cccc|cccc}
  \hline
  \hline
  & $\overline{\ell^{}_{{\rm L}i}}$ & $E^{}_{{\rm R}i}$
  & $N^{}_{{\rm R}i}$ & $S^{}_{{\rm R}i}$
  & $\tilde{H}^{a}_{}$ & $\Phi^{a}_{}$ & $\chi^{a}_{}$ &
  $\phi^{a}_{}$ \\
  \hline
  $q^{}_1$ & 0 & 4 & 2 & 1 & 4 & 3 & 2 & 4\\
  \hline\hline
\end{tabular}
\end{center}
\end{table}

\begin{table}[t]
\caption{The charges of the fermions and scalars under
$Z^{}_{6q^{}_2}$.}
\begin{center}
\begin{tabular}{c|ccc|ccc|ccc|ccc}
  \hline
  \hline
  & $\overline{\ell^{}_{\rm L 1}}$ & $\overline{\ell^{}_{\rm L 2}}$
  & $\overline{\ell^{}_{\rm L 3}}$
  & ~$e^{}_{{\rm R}}$~ & ~$\mu^{}_{{\rm R}}$~ & ~$\tau^{}_{{\rm R}}$~
  & $N^{}_{{\rm R}1}$ & $N^{}_{{\rm R}2}$ & $N^{}_{{\rm R}3}$
  & $S^{}_{{\rm R}1}$ & $S^{}_{{\rm R}2}$ & $S^{}_{{\rm R}3}$ \\
  \hline
  $q^{}_2$ & 0 & 2 & 1 & 0 & 2 & 1 & 0 & 2 & 1 & 0 & 2 & 1\\
  \hline
  \hline
  & $\tilde{H}^{1}_{}$ & $\tilde{H}^{2}_{}$ & $\tilde{H}^{3}_{}$
  & $\Phi^{1}_{}$ & $\Phi^{2}_{}$ & $\Phi^{3}_{}$
  & $\chi^{1}_{}$ & $\chi^{2}_{}$ & $\chi^{3}_{}$
  & $\phi^{1}_{}$ & $\phi^{2}_{}$ & $\phi^{3}_{}$ \\
  \hline
  $q^{}_2$ & 4 & 3 & 2 & 4 & 3 & 2 & 4 & 3 & 2 & 4 & 3 & 2\\
  \hline
  \hline
\end{tabular}
\end{center}
\end{table}

In our model, we choose $n=6$ and $\mathcal{G} = Z^{}_{6
q^{}_1}\times Z^{}_{6 q^{}_2}$. The discrete Abelian group $Z^{}_6$
is given by $Z^{}_6 \equiv \{
1,\omega,\omega^2_{},\omega^3_{},\omega^4_{},\omega^5_{}\}$, where
$\omega=e^{i \pi/3}$. In Tables 1 and 2, we list the charges
$q^{}_1$ and $q^{}_2$ for each field, respectively. The invariance
of the Lagrangian under the $Z^{}_{6 q^{}_1}\times Z^{}_{6q^{}_2}$
leads to the following textures of the Yukawa coupling matrices:
\begin{eqnarray}
Y^{1}_A \sim \left( \begin{matrix} ~0~ &\times & ~0~ \cr \times &
~0~ & ~0~ \cr ~0~ & ~0~ & \times \cr
\end{matrix} \right)\;,\;\;\;\;
Y^{2}_A \sim \left( \begin{matrix} ~0~ & ~0~ & ~0~ \cr ~0~ & ~0~ &
\times \cr ~0~ & \times & ~0~ \cr
\end{matrix} \right)\;\;\;\;{\rm and}\;\;\;\;
Y^{3}_A \sim \left( \begin{matrix} ~0~ & ~0~ & ~0~ \cr 0 & \times &
0 \cr 0 & 0 & 0 \cr
\end{matrix} \right)\;,
\end{eqnarray}
for $Y^{a}_{A}=Y^{a}_{l}$, $Y^{a}_{\rm D}$, $Y^{a}_{\rm S}$,
$Y^{a}_{\rm R}$ and $Y^{a}_{\mu}$. After SSB, the scalars gain their
VEVs, and we are left with the mass terms
\begin{eqnarray}
-\mathcal{L}^{}_{\ell} &=& \overline{E^{}_{\rm L}} M^{}_{l}
E^{}_{\rm R} + \overline{\nu^{}_{\rm L}} M^{}_{\rm D} N^{}_{\rm R} +
\overline{N^{c}_{\rm R}}M^{}_{\rm S} S^{}_{\rm R} + \frac{1}{2}
\overline{N^{c}_{\rm R}} M^{}_{\rm R} N^{}_{\rm R} + \frac{1}{2}
\overline{S^{c}_{\rm R}} M^{}_{\mu} S^{}_{\rm R} +{\rm h.c.} \nonumber\\
&=& \overline{E^{}_{\rm L}} M^{}_{l} E^{}_{\rm R} + \frac{1}{2}
\overline{ \left( \begin{matrix} \nu^{}_{\rm L} & N^{c}_{\rm R} &
S^{c}_{\rm R} \end{matrix} \right) } \left(
\begin{matrix} {\bf 0} & M^{}_{\rm D} & {\bf 0} \cr M^T_{\rm D} &
M^{}_{\rm R} & M^{}_{\rm S} \cr {\bf 0} & M^T_{\rm S} & M^{}_\mu
\end{matrix} \right) \left( \begin{matrix}
\nu^{c}_{\rm L} \cr N^{}_{\rm R} \cr S^{}_{\rm R}
\end{matrix} \right) +{\rm h.c.}\;,
\end{eqnarray}
where $M^{}_{l}$, $M^{}_{\rm D}$, $M^{}_{\rm S}$, $M^{}_{\rm R}$ and
$M^{}_{\mu}$ are mass matrices originating from the Yukawa coupling
matrices and VEVs of the scalars. Taking $M^{}_{\rm D}$ for example,
we arrive at
\begin{eqnarray}
M^{}_{\rm D} &=& \frac{1}{\sqrt{2}} \left[ Y^1_{\rm D}
v(H^1_{})+Y^2_{\rm D} v(H^2_{}) + Y^3_{\rm D} v(H^3_{}) \right] \;,
\end{eqnarray}
in which $v(H^a_{})$ is the VEV of $H^a_{}$. All the mass matrices
$M^{}_{l}$, $M^{}_{\rm D}$, $M^{}_{\rm S}$, $M^{}_{\rm R}$ and
$M^{}_{\mu}$ have the same two-zero textures as
\begin{eqnarray}
\left( \begin{matrix} ~0~ & ~\times~ & ~0~ \cr \times & ~\times~ &
\times \cr ~0~ & ~\times~ & \times \cr
\end{matrix} \right)\;.
\end{eqnarray}
In appendix A, we show that the mass matrix of light left-handed
neutrinos is given by a seesaw-like formula in the physical region:
\begin{eqnarray}
M^{}_{\nu} = - M^{}_{\rm D} \left( M^{}_{\rm R} - M^{}_{\rm S}
M^{-1}_{\mu} M^{T}_{\rm S} \right)^{-1} M^{T}_{\rm D}\;.
\end{eqnarray}
With this formula, one can prove that $M^{}_{\nu}$ also follows the
two-zero texture as in Eq. (8) \cite{FX00,Hu11}. A detailed analysis
will be given in the next section.

We remark that besides $Z^{}_6$, lots of discrete Abelian groups
$Z^{}_n$ can connect the GISM with the UTZT. Even under the same
discrete Abelian group, a different arrangement of the charge
$q^{}_2$ may cause different textures of the Yukawa coupling
matrices $Y^1_{A}$, $Y^2_{A}$ and $Y^3_{A}$, but it keeps the
textures of mass matrices as in Eq. (8) unchanged. In brief, there
are many possibilities to link the GISM with the UTZT. However, if
one requires that the Abalian discrete symmetry be anomaly-free, one
must pay attention to the arrangement for the charges $q^{}_1$ and
$q^{}_2$ of each field to guarantee the anomaly-free conditions
\cite{Anomaly}. Then some arrangements for the charges $q^{}_1$ and
$q^{}_2$ will be ruled out.

\section{The mass texture of active neutrinos}

We have proposed a way to realize the two-zero textures of
$M^{}_{l}$, $M^{}_{\rm D}$, $M^{}_{\rm R}$, $M^{}_{\rm S}$ and
$M^{}_{\mu}$. These textures can be obtained immediately from flavor
symmetries under the direct product of discrete Abelian groups.
However, a realization of the two-zero texture of $M^{}_\nu$ is not
so obvious. To find its texture, we must turn to the matrices
$M^{}_{\rm D}$, $M^{}_{\rm R}$, $M^{}_{\rm S}$ and $M^{}_{\mu}$, all
of which have the same texture zeros. In a way similar to the proof
in Refs. \cite{FX00} and \cite{Hu11}, after giving the two-zero
textures of $M^{}_{\rm D}$, $M^{}_{\rm R}$, $M^{}_{\rm S}$ and
$M^{}_{\mu}$, we can prove that the two-zero textures manifest
themselves again in $M^{}_{\nu}$, as a consequence of Eq. (9). 

We express each matrix $M^{}_a$ (for $a={\rm D}, {\rm S}, {\mu}$) as 
\begin{eqnarray}
M^{}_{a} &=& \left( \begin{matrix} 0 & A^{}_{a} & 0 \cr
A^{}_{a}   & C^{}_{a} & B^{}_{a} \cr 0 & B^{}_{a} &
D^{}_{a} \cr
\end{matrix} \right) \; .
\end{eqnarray}
It is easy to find the inverse matrix of $M^{}_\mu$ has another type of texture zeros
\begin{eqnarray}
M^{-1}_\mu &=& \frac{1}{A^2_{\mu } D^{}_{\mu }}
\left(
\begin{array}{ccc}
 B^{2}_{\mu }-C^{}_{\mu } D^{}_{\mu } & A^{}_{\mu } D^{}_{\mu } & -A^{}_{\mu } B^{}_{\mu } \\
 A^{}_{\mu } D^{}_{\mu } & 0 & 0 \\
 -A^{}_{\mu } B^{}_{\mu } & 0 & A^{2}_{\mu }
\end{array}
\right).
\end{eqnarray}
Then, using the seesaw formula $M^{}_X\equiv - M^{}_{\rm S} M^{-1}_\mu M^{T}_{\rm S}$, we find $M^{}_X$ has the two-zero texture as
\begin{eqnarray}
M^{}_{X} &=& \left( \begin{matrix} 0 & A^{}_{X} & 0 \cr
A^{}_{X}   & C^{}_{X} & B^{}_{X} \cr 0 & B^{}_{X} &
D^{}_{X} \cr
\end{matrix} \right) 
\end{eqnarray}
with 
\begin{eqnarray}
A^{}_X&=& -\frac{A^{2}_{\rm S}}{A^{}_{\mu }}  \;,\nonumber \\
B^{}_X&=& -\frac{A^{}_{\rm S}B^{}_{\rm S}}{A^{}_{\mu }}+\frac{A^{}_{\rm S}B^{}_{\mu } D^{}_{\rm S}}{A^{}_{\mu } D^{}_{\mu }}-\frac{B^{}_{\rm S} D^{}_{\rm S}}{D^{}_{\mu }}  \;,\nonumber \\
C^{}_X&=&  -\frac{2 A^{}_{\rm S} C^{}_{\rm S}}{A^{}_{\mu }}+\frac{A^{2}_{\rm S} C^{}_{\mu }}{A^{2}_{\mu }} 
-\frac{\left(A^{}_{\mu } B^{}_{\rm S}-A^{}_{\rm S} B^{}_{\mu }\right){}^2}{A^{2}_{\mu } D^{}_{\mu }}  \;,\nonumber \\
D^{}_X&=& -\frac{D^{2}_{\rm S}}{D^{}_{\mu }}  \;.
\end{eqnarray}
Thus the two-zero texture is invariant under the seesaw transformation.

Repeating the above process for $M^{}_\nu=-M^{}_{\rm D} (M^{}_{\rm R}+M^{}_{X})^{-1}_{} M^{T}_{\rm D}$, we finally obtain that $M^{}_\nu$ has the two-zero texture as in Eq. (3). The non-zero entries are given by
\begin{eqnarray}
A^{}_\nu&=& -\frac{A^{2}_{\rm D}}{A^{}_{\rm R}+A^{}_X}  \;,\nonumber \\
B^{}_\nu&=& -\frac{A^{}_{\rm D}B^{}_{\rm D}}{A^{}_{\rm R}+A^{}_X}+\frac{A^{}_{\rm D}(B^{}_{\rm R}+B^{}_X) D^{}_{\rm D}}{(A^{}_{\rm R}+A^{}_X) (D^{}_{\rm R}+D^{}_X)}-\frac{B^{}_{\rm D} D^{}_{\rm D}}{D^{}_{\rm R}+D^{}_X}  \;,\nonumber \\
C^{}_\nu&=&  -\frac{2 A^{}_{\rm D} C^{}_{\rm D}}{A^{}_{\rm R}+A^{}_X}+\frac{A^{2}_{\rm D} (C^{}_{\rm R}+C^{}_X)}{(A^{}_{\rm R}+A^{}_X)^2} 
-\frac{\left[(A^{}_{\rm R}+A^{}_X) B^{}_{\rm D}-A^{}_{\rm D} (B^{}_{\rm R}+B^{}_X)\right]{}^2}{(A^{}_{\rm R}+A^{}_X)^2 (D^{}_{\rm R}+D^{}_X)}  \;,\nonumber \\
D^{}_\nu&=& -\frac{D^{2}_{\rm D}}{D^{}_{\rm R}+D^{}_X}  \;.
\end{eqnarray}
It is an exact consequence of the GISM and two-zero textures of $M^{}_{\rm D}$, $M^{}_{\rm R}$, $M^{}_{\rm S}$ and
$M^{}_{\mu}$. 

All the 3 $\times$ 3 mass matrices $M^{}_l$, $M^{}_\nu$, $M^{}_{\rm D}$, $M^{}_{\rm R}$, $M^{}_{\rm S}$ and
$M^{}_{\mu}$ has parallel structures with each other. And they are all fractally similar to the $9\times9$ GISM neutrino matrix $\mathcal{M}$. These similarities can be guaranteed in the framework of flavor symmetries.

\section{Flavor mixing in the UTZT}

In this section we analyze the flavor mixing in the general UTZT case. 
The renormalization-group effect might in general modify the two-zero textures of $M^{}_l$ and $M^{}_\nu$, but it is negligibly small in the inverse seesaw model \cite{Zhang10} because the TeV seesaw scale is so close to the electroweak scale. Hence we just discuss the UTZT at the electroweak scale.

The charged lepton and left-handed neutrino mass matrices with two-zero textures
have been given in Eq. (3), where $A^{}_{l,\nu}$, $B^{}_{l,\nu}$,
$C^{}_{l,\nu}$ and $D^{}_{l,\nu}$ are complex numbers. Some
works on this texture have been done in Refs. \cite{XZH03} and
\cite{More}, but a general analysis has been lacking in the literature.

As a symmetric matrix, $M^{}_l$ can be diagonalized as $M^{}_l =
V^{}_l \hat{M}^{}_l V^T_l$. Here $\hat{M}^{}_l={\rm Diag} \{
m^{}_e,\; m^{}_\mu,\; m^{}_\tau \}$, $V^{}_l=Q^{}_l U^{}_lP^{}_l$,
$Q^{}_l={\rm Diag} \{ e^{i\alpha^{}_l},\; e^{i\beta^{}_l},\; 1 \}$,
$P^{}_l={\rm Diag} \{ e^{i\gamma^{}_e},\; e^{i\gamma^{}_\mu},\;
e^{i\gamma^{}_\tau} \}$ and $U^{}_l$ is given by
\begin{eqnarray}
U^{}_l &=& \left( \begin{matrix} 1 & 0 & 0 \cr 0 & c^{}_e & s^{}_e
\cr 0 & -s^{}_e & c^{}_e \cr
\end{matrix} \right) \left( \begin{matrix} c^{}_\mu & 0 & \hat{s}^{*}_\mu
\cr 0 & 1 & 0 \cr -\hat{s}^{}_\mu & 0 & c^{}_\mu \cr
\end{matrix} \right) \left( \begin{matrix} c^{}_\tau & s^{}_\tau & 0 \cr
-s^{}_\tau & c^{}_\tau & 0 \cr 0 & 0 & 1 \cr
\end{matrix} \right)\; ,
\end{eqnarray}
in which $c^{}_\alpha=\cos\theta^{}_\alpha$,
$s^{}_\alpha=\sin\theta^{}_\alpha$ (for $\alpha= e,\mu,\tau$) and
$\hat{s}^{}_\mu=s^{}_\mu e^{i\delta^{}_\mu}$.

Similarly, $M^{}_\nu$ can be diagonalized as $M^{}_\nu = V^{}_\nu
\hat{M}^{}_\nu V^T_\nu$. Here $\hat{M}^{}_\nu={\rm Diag} \{
m^{}_1,\; m^{}_2,\; m^{}_3 \}$, $V^{}_\nu=Q^{}_\nu U^{}_\nu
P^{}_\nu$, $Q^{}_\nu={\rm Diag} \{ e^{i\alpha^{}_\nu},\;
e^{i\beta^{}_\nu},\; 1 \}$, $P^{}_\nu={\rm Diag} \{
e^{i\gamma^{}_1},\; e^{i\gamma^{}_2},\; e^{i\gamma^{}_3} \}$ and
$U^{}_\nu$ is given by
\begin{eqnarray}
U^{}_\nu &=& \left( \begin{matrix} 1 & 0 & 0 \cr 0 & c^{}_1 & s^{}_1
\cr 0 & -s^{}_1 & c^{}_1 \cr
\end{matrix} \right) \left( \begin{matrix} c^{}_2 & 0 & \hat{s}^{*}_2
\cr 0 & 1 & 0 \cr -\hat{s}^{}_2 & 0 & c^{}_2 \cr
\end{matrix} \right) \left( \begin{matrix} c^{}_3 & s^{}_3 & 0 \cr
-s^{}_3 & c^{}_3 & 0 \cr 0 & 0 & 1 \cr
\end{matrix} \right)\; ,
\end{eqnarray}
in which $c^{}_i=\cos\theta^{}_i$, $s^{}_i=\sin\theta^{}_i$ (for
$i=1,2,3$) and $\hat{s}^{}_2=s^{}_2 e^{i\delta^{}_2}$.

The Maki-Nakagawa-Sakata-Pontecorvo (MNSP) matrix \cite{MNSP} is
defined by $V \equiv V^{\dag}_l V^{}_\nu = P^{\dag}_l U^{\dag}_l
\bar{Q} U^{}_\nu P^{}_\nu$, in which $\bar{Q}={\rm Diag} \{
e^{i\alpha},\; e^{i\beta},\; 1\}$ and $\alpha$, $\beta$ are two
combined parameters defined as $\alpha\equiv
\alpha^{}_\nu-\alpha^{}_l$, $\beta\equiv \beta^{}_\nu - \beta^{}_l$,
respectively. $V$ can be parametrized as $V=QUP$. Here
\begin{eqnarray}
U &=& \left( \begin{matrix} 1 & 0 & 0 \cr 0 & c^{}_{23} & s^{}_{23}
\cr 0 & -s^{}_{23} & c^{}_{23} \cr
\end{matrix} \right) \left( \begin{matrix} c^{}_{13} & 0 &
\hat{s}^{*}_{13} \cr 0 & 1 & 0 \cr -\hat{s}^{}_{13} & 0 & c^{}_{13}
\cr
\end{matrix} \right) \left( \begin{matrix} c^{}_{12} & s^{}_{12} & 0 \cr
-s^{}_{12} & c^{}_{12} & 0 \cr 0 & 0 & 1 \cr
\end{matrix} \right)\; , 
\end{eqnarray}
in which $c^{}_{ij}=\cos\theta^{}_{ij}$,
$s^{}_{ij}=\sin\theta^{}_{ij}$ (for $ij=12,23,13$) and
$\hat{s}^{}_{13} = s^{}_{13}e^{i\delta}$. $P$ and $Q$ are two
diagonal phase matrices. As the charged leptons are the Dirac
fermions, $Q$ is unphysical and can be rotated away by the phase
redefinition of the charged lepton fields. But for the Majorana
neutrinos, only one overall phase in $P$ can be rotated away and the
other two phases are physical. In this case, $P$ can be parametrized
as $P={\rm Diag} \{ e^{i\rho},\; e^{i\sigma},\; 1 \}$.

\subsection{Charged leptons}

Here we derive some relations of the mixing parameters in the
charged lepton sector. Since the (1,1) and (1,3) elements of
$M^{}_l$ are equal to zeros, we obtain
\begin{eqnarray}
\frac{m^{}_e e^{2i\gamma^{}_e}}{m^{}_\tau e^{2i\gamma^{}_\tau}} &=&
- \frac{\hat{s}^{*}_\mu}{c^2_\mu}\left(\frac{c^{}_e
s^{}_\tau}{s^{}_e
c^{}_\tau} +\hat{s}^{*}_\mu \right) \;, \nonumber\\
\frac{m^{}_\mu e^{2i\gamma^{}_\mu}}{m^{}_\tau e^{2i\gamma^{}_\tau}}
&=& + \frac{\hat{s}^{*}_\mu}{c^2_\mu}\left(\frac{c^{}_e
c^{}_\tau}{s^{}_e s^{}_\tau} - \hat{s}^{*}_\mu \right) \;.
\end{eqnarray}
A straightforward calculation leads us to the relations of the
angles
\begin{eqnarray}
\cot^2_{}\theta^{}_e &=& s^2_\mu \left(\sqrt{x^2_l y^2_l
\cot^4_{}\theta_\mu -\sin^2_{}\delta^{}_\mu} -\cos\delta^{}_\mu
\right)\left(\sqrt{ y^2_l \cot^4_{}\theta^{}_\mu
-\sin^2_{}\delta^{}_\mu} +\cos\delta^{}_\mu \right)
 \;, \nonumber\\
\tan^2_{}\theta^{}_\tau &=& \frac{\sqrt{x^2_l y^2_l
\cot^4_{}\theta_\mu -\sin^2_{}\delta^{}_\mu} -\cos\delta^{}_\mu}
{\sqrt{ y^2_l \cot^4_{}\theta^{}_\mu -\sin^2_{}\delta^{}_\mu}
+\cos\delta^{}_\mu }\;,
\end{eqnarray}
and those of the phases
\begin{eqnarray}
\tan (2\gamma^{}_e-2\gamma^{}_\tau+\delta^{}_\mu) &=&
\frac{\sin\delta^{}_\mu} {\sqrt{ x^2_ly^2_l \cot^4_{}\theta^{}_\mu
-\sin^2_{}\delta^{}_\mu}}
\;, \nonumber\\
\gamma^{}_\mu-\gamma^{}_\tau+\delta^{}_\mu/2 &=&0\;,
\end{eqnarray}
where $x^{}_l=m^{}_e/m^{}_\mu$ and $y^{}_l=m^{}_\mu/m^{}_\tau$.
Taking $m^{}_e=0.486 \;{\rm MeV}$, $m^{}_{\mu}=102.7 \;{\rm MeV}$
and $m^{}_{\tau}=1746 \;{\rm MeV}$ at the electroweak scale
\cite{XZZ08} as inputs, we get $x^{}_l = 0.0047$ and $y^{}_l =
0.059$. To assure that Eq. (19) have a real and positive solution,
we require
\begin{eqnarray}
&& 0\leqslant \theta^{}_\mu \leqslant \arctan\sqrt{ x^{}_ly^{}_l }
\approx 1^{\circ}_{}\;,\nonumber\\
&& 0\leqslant \theta^{}_\tau \leqslant
\arctan\sqrt{2 x^{}_l} \approx 6^{\circ}_{}\;, \nonumber\\
&& 0 \leqslant \theta^{}_e \leqslant 90^\circ_{}\;.
\end{eqnarray}
In particular, $\theta^{}_\tau \approx \arctan\sqrt{x^{}_l} \approx
4^\circ_{}$ for $\delta^{}_\mu =\pm 90^\circ_{}$,
$0\leqslant\theta^{}_\tau < 4^\circ_{}$ for $|\delta^{}_\mu| <
90^\circ_{}$, and $4^\circ_{} \leqslant \theta^{}_\tau \leqslant
6^\circ_{}$ for $|\delta^{}_\mu| \geqslant 90^\circ_{}$. Due to the
large mass hierarchy of the charged leptons, $\theta^{}_\mu$ and
$\theta^{}_\tau$ are very small. They can be regarded as the
corrections to the MNSP matrix. Suppressed by $s^{}_\mu$, the phase
$\delta^{}_\mu$ has little influence in the MNSP matrix.
Particularly, we have three special cases:

(1) $\tan\theta^{}_e\ll 1/\sqrt{y^{}_l}$,
\begin{eqnarray}
&&\tan\theta^{}_\mu \approx \sqrt{x^{}_l}\;y^{}_l
\tan\theta^{}_e \;,\nonumber\\
&&\tan\theta^{}_\tau \approx
\sqrt{x^{}_l} \;,\nonumber\\
&&\gamma^{}_e\approx\gamma^{}_\mu \pm 90^\circ_{} \;.
\end{eqnarray}

(2) $\tan\theta^{}_e\gg 1/\sqrt{y^{}_l}$,
\begin{eqnarray}
&&\tan\theta^{}_\mu \approx \sqrt{x^{}_ly^{}_l}
\;,\nonumber\\
&&\tan\theta^{}_\tau \approx \sqrt{\frac{x^{}_l}{y^{}_l}}
\cot\theta^{}_e \;,\nonumber\\
&&\gamma^{}_e\approx\gamma^{}_\mu-\delta^{}_\mu/2 \pm 90^\circ_{}
\;.
\end{eqnarray}

(3) $\tan\theta^{}_e\sim \mathcal{O} \left( 1/\sqrt{y^{}_l}
\right)$, one can find $ \tan\theta^{}_\mu \sim \mathcal{O} \left(
\sqrt{x^{}_ly^{}_l} \right)$ and $\tan\theta^{}_\tau \sim
\mathcal{O} \left( \sqrt{x^{}_l} \right)$ from Eq. (19). In the
leading-order approximation of $s^{}_{\mu}$ and $s^{}_\tau$, we
obtain
\begin{eqnarray}
&& s^2_\mu \approx \frac{-x^{}_ly^2_l
\cos(\gamma^{}_e-\gamma^{}_\mu)} {x^{}_l + y^{}_l
\tan^2_{}\theta^{}_e}
\;,\nonumber\\
&& s^2_\tau \approx \frac{-x^{}_l \cos(\gamma^{}_e-\gamma^{}_\mu)}
{x^{}_l + y^{}_l \tan^2_{}\theta^{}_e}
\;,\nonumber\\
&& \sin\delta^{}_\mu \approx
x^{}_ly^{}_l\sin(\theta^{}_e-\theta^{}_\mu)\;,
\end{eqnarray}
and $\gamma^{}_e-\gamma^{}_\mu$ is arbitrary.

\subsection{Neutrinos}

For the left-handed neutrinos, since the (1,1) and (1,3) elements of
$M^{}_\nu$ equal zeros, we obtain
\begin{eqnarray}
\frac{m^{}_1 e^{2i\gamma^{}_1}_{}}{m^{}_3 e^{2i\gamma^{}_3}_{}} &=&
-\frac{\hat{s}^{*}_2}{c^2_2}\left(\frac{c^{}_1 s^{}_3}{s^{}_1
c^{}_3} +\hat{s}^{*}_2 \right) \;, \nonumber\\
\frac{m^{}_2 e^{2i\gamma^{}_2}_{}}{m^{}_3 e^{2i\gamma^{}_3}_{}} &=&
+\frac{\hat{s}^{*}_2}{c^2_2}\left(\frac{c^{}_1 c^{}_3}{s^{}_1
s^{}_3} - \hat{s}^{*}_2 \right)\;.
\end{eqnarray}
Later in the numerical calculations, we will see that $\theta^{}_2$
is a small angle, in the same order of magnitude as $
\theta^{}_{13}$. In this case, we find $m^{}_1<m^{}_3$ from Eq.
(25). Only the normal hierarchy of neutrino masses is possible in
this texture. A straightforward calculation leads us to the
relations of the angles
\begin{eqnarray}
\cot^2_{}\theta^{}_1 &=& s^2_2 \left(\sqrt{x^2_\nu y^2_\nu
\cot^4_{}\theta^{}_2 -\sin^2_{}\delta^{}_2} -\cos\delta^{}_2 \right)
\left(\sqrt{ y^2_\nu \cot^4_{}\theta^{}_2 -\sin^2_{}\delta^{}_2}
+\cos\delta^{}_2 \right)
 \;, \nonumber\\
\tan^2_{}\theta^{}_3 &=& \frac{\sqrt{x^2_\nu y^2_\nu
\cot^4_{}\theta^{}_2 -\sin^2_{}\delta^{}_2} -\cos\delta^{}_2}
{\sqrt{ y^2_\nu \cot^4_{}\theta^{}_2 -\sin^2_{}\delta^{}_2}
+\cos\delta^{}_2}\;,
\end{eqnarray}
and those of the phases
\begin{eqnarray}
\tan (2\gamma^{}_1-2\gamma^{}_3+\delta^{}_2) &=&
\frac{\sin\delta^{}_2} {\sqrt{x^2_\nu y^2_\nu
\cot^4_{}\theta^{}_2 -\sin^2_{}\delta^{}_2}} \;, \nonumber\\
\tan (2\gamma^{}_2-2\gamma^{}_3+\delta^{}_2) &=&
\frac{-\sin\delta^{}_2} {\sqrt{y^2_\nu \cot^4_{}\theta^{}_2
-\sin^2_{}\delta^{}_2}} \;,
\end{eqnarray}
in which $x^{}_{\nu}=m^{}_1/m^{}_2$, $y^{}_{\nu}=m^{}_2/m^{}_3$ and
$x^{}_{\nu}$, $y^{}_{\nu}<1$. To make Eq. (26) have a real and
positive solution, we require
\begin{eqnarray} && 0 \leqslant \theta^{}_2 \leqslant
\arctan\sqrt{x^{}_\nu
y^{}_\nu}\;,\nonumber\\
&& 0 \leqslant \theta^{}_3 \leqslant \arctan\sqrt{2
x^{}_\nu} \;,\nonumber\\
&& 0 \leqslant \theta^{}_1 \leqslant 90^\circ_{} \;.
\end{eqnarray}

\subsection{The MNSP matrix}

We have obtained some relations of the mixing parameters in both the
charged lepton and left-handed neutrino sectors. Using these
parameters, we can calculate the mixing angles in the MNSP matrix
and some other physical observables. And using the experimental
constraints, we may find the allowed ranges of the parameters and
make predictions for the observables.

The MNSP matrix $V$ can be calculated through $V=V^{\dag}_l
V^{}_\nu$. Considering the smallness of $s^{}_\mu$, $s^{}_\tau$ and
$s^{}_2$, we obtain the approximate expressions of the mixing angles
$\theta^{}_{13}$, $\theta^{}_{12}$ and $\theta^{}_{23}$:
\begin{eqnarray}
\sin\theta^{}_{13}&\approx& \left| \hat{s}^{*}_2 e^{i \alpha}
+c^{}_1 \left(s^{}_e s^{}_{\tau } - c^{}_e \hat{s}^{*}_{\mu }\right)
-s^{}_1 \left(c^{}_e s^{}_{\tau }+ s^{}_e \hat{s}^{*}_{\mu } \right)
e^{i \beta } \right|
\;,\nonumber\\
\tan\theta^{}_{12}&\approx&\left| \tan\theta^{}_3-\frac{e^{-i \alpha
}}{c^2_3} \left[c^{}_1 \left(c^{}_e s^{}_{\tau }+ s^{}_e
\hat{s}^{*}_{\mu } \right) e^{i \beta } +s^{}_1 \left( s^{}_e
s^{}_{\tau } - c^{}_e \hat{s}^{*}_{\mu }\right)\right] \right|
\;,\nonumber\\
\sin\theta^{}_{23}&\approx&\left|c^{}_e s^{}_1e^{i \beta }-c^{}_1
s^{}_e\right| \;.
\end{eqnarray}
These expressions hold to the first order in $s^{}_\mu$, $s^{}_\tau$
and $s^{}_2$. We make some comments on the formulas of the mixing
angles in Eq. (29):
\begin{itemize}
\item Note that $\theta^{}_{13}$ is in the same order of magnitude
as $\theta^{}_2$, and $\theta^{}_2\leqslant\arctan\sqrt{x^{}_\nu
y^{}_\nu}=\arctan\sqrt{m^{}_1/m^{}_3}$. To generate a relatively
large $\theta^{}_{13}$, $m^{}_1$ cannot be too small.
\item Since $\theta^{}_{\mu}$ and $\theta^{}_{\tau}$ are small,
$\theta^{}_{12}\approx\theta^{}_3$ holds. The two-zero texture in
the charged lepton sector just has a small contribution to
$\theta^{}_{12}$.
\item $\theta^{}_{23}$ is an overall result of $\theta^{}_1$,
$\theta^{}_e$ and $\beta$. The two-zero textures in both the charged
lepton and neutrino sectors may have large contributions to
$\theta^{}_{23}$.
\end{itemize}
We conclude that except for $\theta^{}_{12}$, both $\theta^{}_{13}$
and $\theta^{}_{23}$ may receive relatively large corrections from
the charged lepton sector. This is one of the features that make the
UTZT different from the texture zeros discussed in Ref. \cite{GX},
in which $M^{}_l$ is diagonal and only $M^{}_\nu$ has texture zeros.

The strength of CP violation in the neutrino oscillation experiments
is measured by the Jarlskog invariant $\mathcal{J}= {\rm Im}
(V^{}_{e1} V^{}_{\mu2} V^{*}_{e2} V^{*}_{\mu1}) = c^{}_{12}
s^{}_{12} c^{}_{23} s^{}_{23} c^2_{13} s^{}_{13} \sin\delta $
\cite{Jarlskog}. For current experimental data of $\theta^{}_{13}$,
one may expect a relatively large $\mathcal{J}$ if the CP-violating
phase $\delta$ is not suppressed. In the leading-order approximation
of $s^{}_\mu$, $s^{}_\tau$ and $s^{}_2$,
\begin{eqnarray}
\mathcal{J} &\approx& s^{}_\mu J^{}_\mu + s^{}_\tau J^{}_\tau +
s^{}_2J^{}_2 \;,
\end{eqnarray}
in which
\begin{eqnarray}
J^{}_\mu &=& -s^{}_3c^{}_3 \left[s^{}_1 c^{}_e\sin{(\alpha
+\delta^{}_\mu)}-c^{}_1 s^{}_e\sin{(\alpha +\delta^{}_\mu-\beta
)}\right] \left(c^2_1c^2_e+s^2_1s^2_e+2c^{}_1 s^{}_1 c^{}_e
s^{}_e\cos{\beta }\right) \;,\nonumber\\
J^{}_\tau &=& -s^{}_3c^{}_3 \left[ s^{}_1 s^{}_e\sin{\alpha }+c^{}_1
c^{}_e\sin{(\alpha -\beta )} \right]
\left(c^2_1s^2_e+s^2_1c^2_e-2c^{}_1 s^{}_1 c^{}_e  s^{}_e \cos{\beta
}\right) \;, \nonumber\\
J^{}_2 &=& s^{}_3c^{}_3 ( c^{}_e s^{}_e \sin{\beta } \cos{\delta
^{}_2}+c^{}_1 s^{}_1 \sin{\delta^{}_2}\cos{2\theta^{}_e} - c^{}_e
s^{}_e \cos{\beta } \sin{\delta^{}_2}\cos{2\theta^{}_1} ) \;.
\end{eqnarray}
One can see that the first term $s^{}_\mu J^{}_\mu$ is in general
the smallest one because of the smallness of $s^{}_\mu$, and the
last two terms $s^{}_\tau J^{}_\tau$ and $s^{}_2J^{}_2$ may have
comparable contributions to $\mathcal{J}$.

The $0\nu2\beta$ decay experiments are important for examining if
neutrinos are the Majorana fermions. One key parameter in such
experiments is the effective mass $\langle m \rangle^{}_{ee}\equiv
(V \hat{M}^{}_\nu V^{T}_{})^{}_{11} = (V^{\dag}_l M^{}_\nu
V^{*}_l)^{}_{11}$. The pattern in which $M^{}_\nu$ has the two-zero
texture in Eq. (8) and $M^{}_l$ is diagonal gives $\langle m
\rangle^{}_{ee}=0$. Different from such a pattern, the UTZT that we
are considering here yields a non-zero $\langle m \rangle^{}_{ee}$.
In the leading-order approximation of $s^{}_\mu$, $s^{}_\tau$ and
$s^{}_2$, $\langle m \rangle^{}_{ee}$ reads
\begin{eqnarray}
\langle m \rangle^{}_{ee} &\approx& 2 |(U^{}_l)^{*}_{21}
(M^{}_{\nu})^{}_{21}| \;\;\approx\;\; 2 |c^{}_e s^{}_\tau + s^{}_e
\hat{s}^{}_\mu| |A^{}_\nu| \;,
\end{eqnarray}
in which
\begin{eqnarray}
|A^{}_\nu| &\approx& |m^{}_3 s^{}_1 s^{*}_2e^{2i \gamma^{}_3}_{} -
m^{}_1 c^{}_1 c^{}_3 s^{}_3 e^{2i \gamma^{}_1}_{} + m^{}_2 c^{}_1
c^{}_3 s^{}_3 e^{2i \gamma^{}_2}_{}| \;.
\end{eqnarray}
In comparison, the effective mass $\langle m \rangle^{}_{e}$ in the
tritium beta decay is given by
\begin{eqnarray}
\langle m \rangle^{}_{e} &\equiv& \sqrt{(V \hat{M}^{2}_\nu
V^{\dag}_{})^{}_{11}} \;\;\approx\;\; |A^{}_{\nu}| \;.
\end{eqnarray}
Then, we arrive at
\begin{eqnarray}
\frac{\langle m \rangle^{}_{ee}}{\langle m \rangle^{}_{e}} &\approx&
2 |c^{}_e s^{}_\tau + s^{}_e \hat{s}^{}_\mu|\;.
\end{eqnarray}
If we assume $\theta^{}_\tau=4^\circ_{}$ and ignore the smallness of
$\theta^{}_\mu$, then we obtain $\langle m \rangle^{}_{ee}/\langle m
\rangle^{}_{e}\simeq 0.1$.

\subsection{Numerical results}

In the numerical calculations, we choose 7 free parameters
$\theta^{}_e$, $\theta^{}_2$, $\delta^{}_\mu$, $\delta^{}_2$,
$\alpha$, $\beta$ and $m^{}_1$ as inputs. The values of the charged
lepton masses have been given in section 4.2. To be compatible with
the experimental results, we choose $\Delta m^2_{21} \simeq (7.4 -
7.8)\times 10^{-5}_{} \; {\rm eV^2_{}}$, $\Delta m^2_{31} \simeq
(2.4 - 2.7) \times 10^{-3}_{}\; {\rm eV^2_{}}$,
$\theta^{}_{23}\simeq (42^\circ_{}-49^\circ_{})$,
$\theta^{}_{12}\simeq (33^\circ_{}-35^\circ_{})$ and the new data
$\theta^{}_{13}\simeq (8.0^\circ_{}-9.6^\circ_{})$ from the Daya Bay
experiment as constraints. With the help of these data, we can
obtain the allowed ranges of the input parameters and calculate the
observables.

In Fig. 1, we show the comparison of the values between
$\theta^{}_1$ and $\theta^{}_e$ and that of the values between
$\theta^{}_2$ and $\theta^{}_\tau$. The first two angles are
dominant parameters in the expression of $\theta^{}_{23}$, while the
last two angles are dominant parameters in the expression of
$\theta^{}_{13}$ (see Eq. (29)). Numerically, we obtain
$\theta^{}_1\simeq(24^\circ_{} - 72^\circ_{})$ versus
$\theta^{}_e\simeq(0 - 90^\circ_{})$ for $\theta^{}_{23}$, and
$\theta^{}_2\simeq(4^\circ_{} - 13^\circ_{})$ versus
$\theta^{}_\tau\simeq(0 - 6^\circ_{})$ for $\theta^{}_{13}$. A lot
of points are located around $\theta^{}_\tau= 4^\circ_{}$,
indicating that $\delta^{}_\mu\approx \pm 90^\circ_{}$ is favored.

In Fig. 2, we show the parameter space and some phenomenological
predictions in the general case. We plot the allowed regions of
$(\theta^{}_{13},m^{}_1)$ and $(\theta^{}_{12},\theta^{}_{23})$
parameters first in the figure. The lightest neutrino mass $m^{}_1$
is constrained in the range $(0.001 - 0.015)\; {\rm eV}$. The points
of the mixing angles $\theta^{}_{12}$, $\theta^{}_{23}$ and
$\theta^{}_{13}$ are nearly evenly distributed in the full parameter
space. Predictions for parameters related to CP violation are shown
then. There is little restriction on the combined input parameters
$\alpha$ and $\beta$ except that $\beta$ is more likely to approach
$\pm90^\circ_{}$. For the Majorana phases $\rho$ and $\sigma$, the
relation $\rho\simeq \sigma \pm90^\circ_{}$ holds roughly. The
numerical result of the Jarlskog invariant $\mathcal{J}$ is also
shown in Fig. 2. Due to the largeness of $\theta^{}_{13}$,
$|\mathcal{J}|$ can reach several percent. Concretely, it can
maximally reach $0.03$ at $\theta^{}_{13}=8^\circ_{}$ and $0.04$ at
$\theta^{}_{13}=9.6^\circ_{}$. The effective masses in the tritium
beta decay and $0\nu2\beta$ decay are shown at the end of Fig. 2.
One can see that the ratio $\langle m \rangle^{}_{ee} / \langle m
\rangle^{}_{e}$ is of $\mathcal{O} (0.1)$ in most cases. Since
$\langle m \rangle^{}_{e}\simeq 0.01\; {\rm eV}$ is referred in Fig.
2, $\langle m \rangle^{}_{ee}$ can maximally reach $ 10^{-3}\; {\rm
eV}$. However, this is still below the sensitivity of the near
future experiments, which is expected to be $\langle m
\rangle^{}_{ee} \simeq (1-5)\times10^{-2} \; {\rm eV}$
\cite{Giunti12}.

One can reconstruct the charged lepton and left-handed neutrino mass
matrices with the help of the experimental constraints. Considering
that there are cancelations in some special cases, leading to
vanishing values of $A^{}_{l,\nu}$, $B^{}_{l,\nu}$, $C^{}_{l,\nu}$
or $D^{}_{l,\nu}$, the positive lower bounds may not exist. But one
can expect that there are some ranges in which most of the points
are located. In our calculation, we find that 95\% of the points are
located in the following ranges:
\begin{eqnarray}
&&|A^{}_l| \simeq {\rm(7.4 - 31)} \;{\rm MeV} \;,\;\;\;\;\;
|B^{}_l| \simeq {\rm(0.046 - 0.94)} \;{\rm GeV} \;,\nonumber\\
&&|C^{}_l| \simeq {\rm(0.96 - 1.8)} \;{\rm GeV} \;,\;\;\; |D^{}_l|
\simeq {\rm(0.11 - 1.8)} \;{\rm GeV} \;,
\end{eqnarray}
and
\begin{eqnarray}
&&|A^{}_\nu| \simeq {\rm(0.0073 - 0.018)} \;{\rm eV} \;,\;\;
|B^{}_\nu| \simeq {\rm(0.019 - 0.028)} \;{\rm eV} \;,\nonumber\\
&&|C^{}_\nu| \simeq {\rm(0.011 - 0.040)} \;{\rm eV} \;,
\;\;\;\;|D^{}_\nu| \simeq {\rm(0.010 - 0.048)} \;{\rm eV} \;.
\end{eqnarray}
In the neutrino sector, all the elements of $M^{}_{\nu}$ are in the
$\mathcal{O}(0.01)$ eV order. But in the charged lepton sector, the
elements of $M^{}_{l}$ vary within some wide ranges because of the
uncertainty of $\theta^{}_e$.

\section{Large $\bf\theta^{}_{13}$ and two ans$\ddot{\bf a}$tze
of the UTZT}

In the previous section, we have considered the UTZT in the general
case. Since there are 7 free parameters as inputs, it does not get
stringent experimental constraints. We shall consider some special
cases of the UTZT to simplify its texture.

First, we assume that the condition \cite{XZH03}
\begin{equation}
\arg (C^{}_{l,\nu}) + \arg (D^{}_{l,\nu})  = 2 \arg (B^{}_{l,\nu}) \;
\end{equation}
is satisfied. Then $M^{}_{l}$ and $M^{}_{\nu}$ are respectively
decomposed into
\begin{eqnarray}
M^{}_l  = P^T_l \overline{M}^{}_l P^{}_l e^{2i\gamma^{}_\tau} &{\rm
and}&  M^{}_\nu = P^T_\nu \overline{M}^{}_\nu P^{}_\nu
e^{2i\gamma^{}_3} \; ,
\end{eqnarray}
in which
\begin{eqnarray}
\overline{M}^{}_{l,\nu}  &=&  \left( \begin{matrix} 0 &
|A^{}_{l,\nu}| & 0 \cr |A^{}_{l,\nu}|   & |C^{}_{l,\nu}| &
|B^{}_{l,\nu}| \cr 0 & |B^{}_{l,\nu}| & |D^{}_{l,\nu}| \cr
\end{matrix} \right) \;.
\end{eqnarray}
In the following discussions, we turn to two different ans$\rm
\ddot{a}$tze: ansatz (A), $|A^{}_{l,\nu}|=|D^{}_{l,\nu}|$; and
ansatz (B), $|C^{}_{l}|=|B^{}_{l}|$ and $|C^{}_{\nu}|=|D^{}_{\nu}|$.

\subsection{Ansatz (A)}

We propose to consider this new ansatz, in which both
$|A^{}_{l}|=|D^{}_{l}|$ and $|A^{}_{\nu}|=|D^{}_{\nu}|$ hold.  
Our motivations are based on the model which we built in section 2:
\begin{itemize}
\item
In the charged lepton sector, the (1,2) and (3,3) entries of the
Yukawa coupling matrix $Y^1_{l}$ are nonzero. It is natural to assume
that they have the same magnitude: $|(Y^1_{l})^{}_{12}| =
|(Y^1_{l})^{}_{33}|$. After SSB, we arrive at $|(M^{}_{l})^{}_{12}|
= |(M^{}_{l})^{}_{33}|$, or equivalently, $|A^{}_{l}| = |D^{}_{l}|$.
In this assumption, we can reduce the number of free input parameters.
This equality can be realized in the non-Abelian discrete group $A^{}_5$ \cite{Ding} with suitable arrangements of the particle contents\footnote{
We may arrange $\ell^{}_{\rm L}$ and $E^{}_{\rm R}$ as the triplets, $H^1$ as a singlet, and embed $H^2_{}$ and $H^3_{}$ to a 5-plet in $A^{}_5$. After $H^1$ gains its vacuum expectation value, we are led to $A^{}_l=D^{}_l$. With a suitable vacuum alignment for the 5-plet, no additional mass term will be introduced and the two-zero texture is preserved. 
}.

\item
Applying the above discussion to the neutrino sector, we
are led to
\begin{eqnarray}
|(M^{}_{\rm D})^{}_{12}| = |(M^{}_{\rm D})^{}_{33}|\;, & |(M^{}_{\rm
S})^{}_{12}| = |(M^{}_{\rm S})^{}_{33}|\;,\nonumber\\
|(M^{}_{\rm R})^{}_{12}| = |(M^{}_{\rm R})^{}_{33}|\;, &
|(M^{}_{\mu})^{}_{12}| = |(M^{}_{\mu})^{}_{33}|\;.
\end{eqnarray}
Then, using the inverse seesaw formula in Eq. (9), we arrive at
\begin{eqnarray}
|(M^{}_{\nu})^{}_{12}| = |(M^{}_{\nu})^{}_{33}| = \frac{|(M^{}_{\rm
D})^{}_{12}|^2_{} |(M^{}_{\mu})^{}_{12}|}{|(M^{}_{\rm
S})^{}_{12}|^2_{}- |(M^{}_{\mu})^{}_{12}| |(M^{}_{\rm R})^{}_{12}|}
\;,
\end{eqnarray}
or equivalently, $|A^{}_{\nu}| = |D^{}_{\nu}|$.
\end{itemize}

In ansatz (A), the mass matrices in both the charged lepton and
left-handed neutrino sectors can be solved exactly in terms of their
mass eigenvalues. In the left-handed neutrino sector, we have the
expression of $\overline{M}^{}_{\nu}$ in terms of its three mass
eigenvalues
\begin{eqnarray}
|A^{}_{\nu}| & = & \left( m^{}_1 m^{}_2 m^{}_3 \right) ^{1/3} \; ,
\nonumber \\
|B^{}_{\nu}| & = & \large[\left( m^{}_1 m^{}_2 m^{}_3 \right) ^{1/3}
\left( m^{}_1-m^{}_2+m^{}_3 \right) - 2 \left( m^{}_1 m^{}_2 m^{}_3
\right) ^{2/3} \large
\nonumber\\
&&\large +m^{}_1m^{}_2 - m^{}_1m^{}_3 +m^{}_2m^{}_3 \large]^{1/2}\;
,
\nonumber \\
|C^{}_{\nu}| & = & m^{}_1-m^{}_2+m^{}_3-\left( m^{}_1 m^{}_2 m^{}_3
\right) ^{1/3} \; ,
\end{eqnarray}
and that of $U^{}_{\nu}$ in terms of the ratios of the eigenvalues
\begin{eqnarray}
U^{}_{\nu} &=& \left( \begin{matrix} k^{}_{\nu1} (x^{}_{\nu}
y^{}_{\nu}-a^{}_{\nu}) a^{}_{\nu} & k^{}_{\nu2}
(y^{}_{\nu}+a^{}_{\nu}) a^{}_{\nu} & k^{}_{\nu3} (1-a^{}_{\nu})
a^{}_{\nu} \cr k^{}_{\nu1} (x^{}_{\nu} y^{}_{\nu}-a^{}_{\nu})
x^{}_{\nu} y^{}_{\nu} & -k^{}_{\nu2} (y^{}_{\nu}+a^{}_{\nu})
y^{}_{\nu} & k^{}_{\nu3}(1-a^{}_{\nu}) \cr k^{}_{\nu1} b^{}_{\nu}
x^{}_{\nu}y^{}_{\nu} & k^{}_{\nu2}b^{}_{\nu} y^{}_{\nu} &
k^{}_{\nu3}b^{}_{\nu} \cr
\end{matrix} \right) \;,
\end{eqnarray}
where $a^{}_{\nu}=|A^{}_{\nu}|/m^{}_3$,
$b^{}_{\nu}=|B^{}_{\nu}|/m^{}_3$, $c^{}_{\nu}=|C^{}_{\nu}|/m^{}_3$
and
\begin{eqnarray}
k^{}_{\nu1} &=& \left[ \left(a^2_{\nu}+x^2_{\nu}y^2_{\nu} \right)
(x^{}_{\nu}y^{}_{\nu}-a^{}_{\nu})^2
+x^2_{\nu}y^2_{\nu} b^2_{\nu} \right]^{-1/2} \;,\nonumber\\
k^{}_{\nu2} &=& \left[(a^2_{\nu}+y^2_{\nu})
(y^{}_{\nu}+a)^2+y^2_{\nu} b^2_{\nu}\right]^{-1/2}\;,\nonumber\\
k^{}_{\nu3} &=& \left[(a^2_{\nu}+1)(1-a^{}_{\nu})^2
+b^2_{\nu}\right]^{-1/2}\;.
\end{eqnarray}

In the charged lepton sector, after replacing the index $\nu \to l$
and the masses $(m^{}_1,m^{}_2,m^{}_3) \to
(m^{}_e,m^{}_\mu,m^{}_\tau)$, we arrive at the expressions of
$\overline{M}^{}_{l}$ and $U^{}_{l}$. The relations
\begin{eqnarray}
\gamma^{}_{e,1} = \gamma^{}_{\mu,2} \pm 90^\circ_{} =
\gamma^{}_{\tau,3}\;
\end{eqnarray}
must be required in the phase matrices $P^{}_{l,\nu}$, while
$Q^{}_{l,\nu}$ are arbitrary.

The mixing angles $\theta^{}_{12}$, $\theta^{}_{23}$,
$\theta^{}_{13}$ and the CP phases $\delta$, $\rho$, $\sigma$ can be
obtained from $V \equiv V^{\dag}_l V^{}_\nu = P^{\dag}_l U^{\dag}_l
\bar{Q} U^{}_\nu P^{}_\nu$.

The numerical results of the parameter space and phenomenological
predictions in ansatz (A) are shown in Fig. 3. Only 3 free
parameters, $m^{}_1$, $\alpha$ and $\beta$, are taken as inputs. The
experimental constraints are the same as those in the general case.
The constraint on $m^{}_1$ in this ansatz is much stronger than that
in the general case. One can get $m^{}_1\simeq(0.002 - 0.003)\; {\rm
eV}$ in Fig. 3. Although the number of free parameters has decreased
to 3, the numerical results of the mixing angles $\theta^{}_{12}$,
$\theta^{}_{23}$ and $\theta^{}_{13}$ still fit the experimental
constraints very well. Among them, $\theta^{}_{12}$ and
$\theta^{}_{13}$ are still nearly evenly distributed in the
parameter space, and $\theta^{}_{23}$ has a very slight preference
for being larger than $45^\circ_{}$. The CP-violating parameters are
constrained more stringently. The allowed region of the
$(\alpha,\beta)$ parameters is much smaller:
$|\alpha|\simeq(45^\circ_{}-90^\circ_{})$ and
$|\beta|\simeq(120^\circ_{}-180^\circ_{})$. $|\mathcal{J}|$ can
maximally reach 0.02 at $\theta^{}_{13}=8^\circ_{}$ and 0.03 at
$\theta^{}_{13}=9.6^\circ_{}$. The relation
$\rho\approx\sigma\pm90^\circ_{}$ is a good approximation. The ratio
$\langle m \rangle^{}_{ee}/\langle m \rangle^{}_{e}$ is more likely
to get a small value than that in the general case. It is only
allowed in the range $(0.002-0.04)$. Taking $\langle m
\rangle^{}_{e} \simeq 10^{-2}\; {\rm eV}$, we obtain $\langle m
\rangle^{}_{ee} \simeq (0.2-4)\times 10^{-4}\; {\rm eV}$. This is
far beyond the sensitivity of the future experiments.

\subsection{Ansatz (B)}

In ansatz (B), the requirements $|C^{}_{l}|=|B^{}_{l}|$ and
$|C^{}_{\nu}|=|D^{}_{\nu}|$ are imposed. This ansatz was first
proposed in Ref. \cite{XZH03}. It is motivated by the mass hierarchy
of the charged leptons and the experimental fact that the mixing
angle $\theta^{}_{23}$ in the MNSP matrix is about $45^\circ_{}$.
The relation $|C^{}_l|=|B^{}_l|$ will lead to $|C^{}_l|\approx|m_\mu|$, which
is compatible with the fact that charged leptons have a large mass
hierarchy. And the requirement $|C^{}_\nu|=|D^{}_\nu|$ can lead to 
$\theta_{23} = 45^\circ$ easily. A detailed interpretation for
this ansatz can be found therein. Here we reanalyze it by using the
latest experimental data.

The solutions for diagonalizing $M^{}_l$ and $M^{}_\nu$ in terms of
the mass eigenvalues and their ratios have been give in Ref.
\cite{XZH03}. We use them for our numerical calculation and show the relevant results
 in Fig. 4. 
The same inputs and constraints in ansatz (A) are applied to this ansatz. The
lightest neutrino mass $m^{}_1$ is given by $m^{}_1\simeq(0.004 -
0.008)\; {\rm eV}$, bigger than that in ansatz (A). For the mixing
angles, $\theta^{}_{13}>8.8^\circ_{}$ and
$\theta^{}_{12}<33.8^\circ_{}$ hold, and $\theta^{}_{23}$ is easier
to gain a value smaller than $45^\circ_{}$. As shown in Fig. 4, two
thirds of the $(\theta^{}_{12}, \theta^{}_{23})$ parameter space is
excluded. The constraint on the $(\alpha,\beta)$ parameter space is
still loose and the relation $\rho\approx\sigma\pm90^\circ_{}$ is
also valid. $|\mathcal{J}|$ in this ansatz can maximally reach
$0.02$ at $\theta^{}_{13}=9.6^\circ_{}$, smaller than the maximal
value in ansatz (A). The prediction for the effective mass of the
$0\nu2\beta$ decay is totally different from that in ansatz (A). It
gives $\langle m \rangle^{}_{ee}/\langle m \rangle^{}_{e}\simeq
0.1$. Since $\langle m \rangle^{}_{e} \simeq 0.01\; {\rm eV}$ also
holds in this ansatz, we arrive at $\langle m \rangle^{}_{ee} \simeq
0.001\; {\rm eV}$. We can compare the new results with the old ones
presented in Ref. \cite{XZH03}. Since the mixing parameters are
measured more precisely, most part of the parameter space is
excluded. Ansatz (B) now is not so favored as before.

In this section, we have analyzed the UTZT in two ans$\ddot{\rm
a}$tze. They have two main different features distinguishing
themselves from each other. One is the difference of the parameter
space of the mixing angles. Ansatz (A) is favored in the full
$(\theta^{}_{12},\theta^{}_{23})$ parameter space, while ansatz (B)
is just partly favored. This feature makes ansatz (A), which is a
natural assumption of our model in section 2, more interesting than
ansatz (B). The other feature is the prediction for $\langle m
\rangle^{}_{ee}$. The value of $\langle m \rangle^{}_{ee}$ in ansatz
(B) is much larger than that in ansatz (A), although both are below
the sensitivity of the near future experiments.

\section{Conclusion}
The GISM gives vanishing (1,1) and (1,3) submatrices of the
$9\times9$ neutrino mass matrix $\mathcal{M}$. This is similar to
the UTZT which gives vanishing (1,1) and (1,3) elements of the
$3\times3$ mass matrices $M^{}_{l,\nu}$. We have pointed out their
similarity and considered their several aspects. The main points are
listed in the following.

(1) We have proposed a model based on the discrete Abelian group
$Z^{}_6\times Z^{}_6$ to realize both the GISM and the UTZT. We
reiterate that besides $Z^{}_6$ there are many discrete Abelian
groups whose direct products can realize both of them.


(2) We have calculated the UTZT in the general case. Only the normal
hierarchy of the neutrino masses is allowed by this texture. We
obtain the lightest neutrino mass $m^{}_1 \simeq (0.001-0.015)$ eV.
The Jarlskog invariant $\mathcal{J}$ can maximally reach 0.04 in
view of the new experimental results of $\theta^{}_{13}$. The
effective mass $\langle m\rangle^{}_{ee}$ in the $0\nu2\beta$ decay
can maximally reach 0.001 eV.

(3) We have compared two ans$\ddot{\rm a}$tze of the UTZT. Ansatz
(A) is a natural approximation of our model built in section 2, and
ansatz (B) is a special case which has been considered in Ref.
\cite{XZH03}. The mixing angles in ansatz (A) fit the experimental
constraints quite well, while in ansatz (B),
$\theta^{}_{13}>8.8^\circ_{}$ and $\theta^{}_{12}<33.8^\circ_{}$ are
allowed, and $\theta^{}_{23}<45^\circ_{}$ is preferred. Ansatz (B)
predicts the effective mass $\langle m \rangle^{}_{ee}\simeq {\rm
0.001\; eV}$ in the $0\nu2\beta$ decay experiments, while ansatz (A)
can only predict $\langle m \rangle^{}_{ee}$ one or two orders of
magnitude smaller than that in ansatz (B).

Finally, we stress that the GISM can avoid the hierarchy problem and
is testable in collider experiments, and the UTZT agrees very well
with current neutrino oscillation data. Both the GISM and UTZT can
be realized from the same Abelian symmetry due to their similar
structures, 
although their uniqueness connot easily be verified in the bottom-up approach of model building. 
Except for the above discussions, there are some other
interesting aspects of the GISM and UTZT in neutrino phenomenology.
One is to discuss possible collider signatures of the TeV-scale right-handed or
additional gauge-singlet neutrinos in the GISM, which could be
explored by the Large Hadron Collider. Another aspect is related to
the baryogenesis via leptogenesis, so as to account for the
cosmological matter-antimatter asymmetry. The two-zero textures of
the Yukawa coupling matrices and the uncertainty of the scales of
$M^{}_{\rm R}$ and $M^{}_{\mu}$ may affect how the leptogenesis
mechanism works in the early Universe. A detailed analysis of these
aspects will be done elsewhere.

\subsection*{Acknowledgments}

The author would like to thank Prof. Z.Z. Xing for suggesting this
work and correcting the manuscript in great detail. He is also
grateful to T. Araki, G.J. Ding, Y.F. Li and H. Zhang for useful discussions and good
advices. This work was supported in part by the National Natural
Science Foundation of China under grant No. 11135009.

\appendix

\section{Simplification of the neutrino mass matrix in the GISM}

\subsection{General analysis}

The neutrino mass matrix in the GISM is described by a $9\times9$
matrix $\mathcal{M}$ given in Eq. (2), where $M^{}_{\rm D}$,
$M^{}_{\rm S}$, $M^{}_{\rm R}$ and $M^{}_\mu$ are $3\times3$ complex
submatrices. For physical conditions, one can naturally assume that
the scale of $M^{}_{\rm D}$ is the electroweak scale and the scale
of $M^{}_{\rm S}$ is several orders larger than that of $M^{}_{\rm
D}$. To some extent, the scales of $M^{}_{\rm R}$ and $M^{}_{\mu}$
are more arbitrary. They can be either very high or very small due
to different mechanisms. Large mass scales can be regarded as the
breaking of a certain symmetry at a very high energy scale, similar
to the Majorana mass matrix of the right-handed neurinos in the
type-I seesaw model. And small mass scales may be generated from
higher dimensional operator after integrating out some unknown heavy
fields \cite{Ma09c}. Small mass scales are also consistent with the
't Hooft's naturalness criterion \cite{Hooft}, because the
conservation of the lepton number is recovered when $M^{}_{\rm R}$
and $M^{}_{\mu}$ reduce to zeros.

Since different scales of $M^{}_{\rm R}$ and $M^{}_{\mu}$ may lead
to different phenomenological consequences, it is necessary to do a
general analysis of how $\mathcal{M}$ can be simplified in different
cases. We denote
\begin{eqnarray}
M'_{\rm D} = \left( \begin{matrix} M^{}_{\rm D} & {\bf 0}
\end{matrix} \right) \;\;\;\; {\rm and} \;\;\;\;
M'_{\rm R} = \left( \begin{matrix} M^{}_{\rm R} & M^{}_{\rm S} \cr
M^T_{\rm S} & M^{}_\mu \end{matrix} \right) \;.
\end{eqnarray}
Obviously, the scale of $M'_{\rm D}$ is several orders smaller than
that of $M'_{\rm S}$, and $M'_{\rm R}$ yields the masses of the
right-handed and additional gauge-singlet neutrinos. One can obtain
the mass matrix of light left-handed neutrinos through a seesaw-like
formula
\begin{eqnarray}
M^{}_{\nu} &\approx& -M'_{\rm D} {M'_{\rm R}}^{-1}_{} {M'_{\rm
D}}^T_{} \;\;=\;\; - M^{}_{\rm D} \left( M^{}_{\rm R} - M^{}_{\rm S}
M^{-1}_{\mu} M^{T}_{\rm S} \right)^{-1} M^{T}_{\rm D}\;.
\end{eqnarray}
The mass formula in Eq. (48) is the main result in the GISM. It can
be further simplified in some special cases. However, since Eq. (48)
is only valid for $M^{}_{\rm D} \ll M^{}_{\rm R} - M^{}_{\rm S}
M^{-1}_{\mu} M^{T}_{\rm S}$, the exception should also be considered
especially.

\subsection{Special cases}

For different mass scales of $M^{}_{\rm R}$ and $M^{}_{\mu}$, the
expression of $M^{}_{\nu}$ in Eq. (48) can be simplified. For the
sake of convenience in the following dicussions, we denote $M'_{\rm
R}$ to be block-diagonalized by a $6\times6$ unitary matrix $W$ as
\begin{eqnarray}
W^{\dag}_{}M'_{\rm R}W^{*}_{} &\equiv&   \left(
\begin{matrix} M^{}_{\rm m} & {\bf 0} \cr {\bf 0} &
M^{}_{\rm h} \end{matrix} \right) \;,
\end{eqnarray}
in which $M^{}_{\rm m}$ and $M^{}_{\rm h}$ are $3\times3$ matrices
standing for the medium and heaviest neutrino masses, respectively.
Here we consider three typical cases to simplify the mass matrices
$M^{}_{\nu}$ and $M'_{\rm R}$.

{\bf Case (A):} ${\bf 0} \leqslant M^{}_{\rm R}\ll M^{}_{\rm S}$ and
${\bf 0} \leqslant M^{}_{\mu} \ll M^{}_{\rm S}$. Eq. (48) can be
simplified to \cite{Parida}
\begin{eqnarray}
M^{}_{\nu} &\approx& M^{}_{\rm D} M^{-T}_{\rm S} M^{}_{\mu}
M^{-1}_{\rm S} M^{T}_{\rm D}\;,
\end{eqnarray}
and $M'_{\rm R}$ is simplified to
\begin{eqnarray}
M'_{\rm R} &\approx& \left( \begin{matrix} M^{}_{\rm R} & M^{}_{\rm S}
\cr M^T_{\rm S} & M^{}_\mu
\end{matrix} \right)\;.
\end{eqnarray}
This case has been discussed in Ref. \cite{Ma09b}.
Since $M^{}_\mu$ and $M^{}_{\rm R}$ are much smaller than $M^{}_{\rm S}$, the right-handed and additional gauge-singlet neutrinos have nearly
degenerate masses and are combined to form the pseudo-Dirac
particles.  Their masses can be not huge and may be testable by the collider.
For instance, assuming $M^{}_{\nu}\sim {\rm 0.1\; eV}$, $M^{}_{\mu} \sim
{\rm 1\; keV}$ and $M^{}_{\rm D} \sim {\rm 10\; GeV}$, we obtain
$M^{}_{\rm S} \sim {\rm 1\; TeV}$. 
Another aspect of this case is the non-unitary effects. Such effects in the mixing matrix are approximate to $M^{}_{\rm D} M^{-1}_{\rm S}$. Experimental data show that they are smaller than $\mathcal{O}(1)$ \cite{Antusch}.
Due to present accuracies for measuring mixing angles, we do not have to consider the non-unitary effects in the mixing matrix.
We will ignore them in the main body of this paper.

{\bf Case (B):} $M^{}_{\rm \mu} \ll M^{}_{\rm S} \ll M^{}_{\rm R}$.
$M'_{\rm R}$ can be simplified to
\begin{eqnarray}
M^{}_{\rm m} &\approx& M^{}_{\mu} - M^{T}_{\rm S} M^{-1}_{\rm R}
M^{}_{\rm
S} \;, \nonumber\\
M^{}_{\rm h} &\approx& M^{}_{\rm R} \;.
\end{eqnarray}
This case accommodates a large range of the masses of sterile
neutrinos and provides a possibility for low scale leptogenesis
\cite{Kim07}. One can further discuss the case (B1): $M^{}_{\rm R}
\ll M^{}_{\rm S} M^{-1}_{\mu} M^{T}_{\rm S}$ and case (B2):
$M^{}_{\rm R} \gg M^{}_{\rm S} M^{-1}_{\mu} M^{T}_{\rm S}$. In case
(B1), $M^{}_{\nu}$ can be simplified to Eq. (50); while in case
(B2), $M^{}_{\nu}$ can be simplified to \cite{Kim07}
\begin{eqnarray}
M^{}_{\nu} &\approx& - M^{}_{\rm D} M^{-1}_{\rm R} M^{T}_{\rm D}\;.
\end{eqnarray}
To derive the tiny left-handed neutrino masses in case (B2), the
scale of $M^{}_{\rm R}$ should in general be very high, which is
similar to the type-I seesaw model.

{\bf Case (C):} $M^{}_{\rm R}\gg M^{}_{\rm S}$ and $M^{}_{\mu} \gg
M^{}_{\rm S}$. In this case we obtain Eq. (53) and
\begin{eqnarray}
M^{}_{\rm m} &\approx& \max (M^{}_{\rm R},M^{}_{\mu})\;,
\nonumber\\
M^{}_{\rm h} &\approx& \min (M^{}_{\rm R},M^{}_{\mu})\;.
\end{eqnarray}
The choice of the mass scale of $M^{}_{\mu}$ is a little arbitrary
except for $M^{}_{\mu} \gg M^{}_{\rm S}$. There is only small mixing
between the right-handed and additional gauge-singlet neutrinos.

Since Eqs. (50) and (53) are the typical formulas of left-handed
neutrino mass matrices in the OISM and type-I seesaw model,
respectively, these two models can be regarded as two special cases
of the GISM to some extent.

\subsection{Exception}

Note that Eq. (48) does not hold for $M^{}_{\rm R} - M^{}_{\rm S}
M^{-1}_{\mu} M^{T}_{\rm S}$ $ \lesssim M^{}_{\rm D}$. This exception
should be considered in particular. It can be further divided into
two cases: case (D), $M^{}_{\rm R}\lesssim M^{}_{\rm D}$ and
$M^{}_{\rm S} M^{-1}_{\mu} M^{T}_{\rm S}$ $ \lesssim M^{}_{\rm D}$;
and case (E), $M^{}_{\rm R} \gg M^{}_{\rm D}$ and $M^{}_{\rm S}
M^{-1}_{\mu} M^{T}_{\rm S} $ $ \gg M^{}_{\rm D}$, but there is a
cancelation that leads to $M^{}_{\rm R}-M^{}_{\rm S} M^{-1}_{\mu}
M^{T}_{\rm S} \lesssim M^{}_{\rm D}$.

{\bf Case (D).} Since $M^{}_{\rm S}$ is several orders higher than
$M^{}_{\rm D}$, we are led to $M^{}_{\rm R} \ll M^{}_{\rm S} \ll
M^{}_{\mu}$. $\mathcal{M}$ can be simplified by a congruent
transformation with a $9\times9$ unitary matrix $\mathcal{W}$. One
can write out $\mathcal{W}$ and
$\mathcal{W}^{\dag}_{}\mathcal{M}\mathcal{W}^{*}_{}$ as
\begin{eqnarray}
\mathcal{W} &\approx& \left(
\begin{matrix} {\bf 1} & {\bf 0} & {\bf 0} \cr {\bf 0} &
{\bf 1} & -M^{}_{\rm S}M^{-1}_{\mu} \cr {\bf 0} & \left(M^{}_{\rm
S}M^{-1}_{\mu}\right)^{\dag}_{} & {\bf 1}
\end{matrix} \right) \;,\nonumber\\
\mathcal{W}^{\dag}_{}\mathcal{M}\mathcal{W}^{*}_{} &\approx& \left(
\begin{matrix} {\bf 0} & M^{}_{\rm D} & {\bf 0} \cr M^{T}_{\rm D} &
M^{}_{\rm R} - M^{T}_{\rm S} M^{-1}_{\rm R} M^{}_{\rm S} & {\bf 0}
\cr {\bf 0} & {\bf 0} & M^{}_{\mu}
\end{matrix} \right) \;,
\end{eqnarray}
respectively. Finally, we obtain $M^{}_{\nu} \approx M^{}_{\rm D}$,
which is too heavy to be the left-handed neutrino mass matrix. Thus,
this case is not interesting.

{\bf Case (E).} $\mathcal{W}$ and
$\mathcal{W}^{\dag}_{}\mathcal{M}\mathcal{W}^{*}_{}$ are given by
\begin{eqnarray}
\mathcal{W} &\approx& \left( \begin{matrix} {\bf 1} & {\bf 0} & {\bf
0} \cr {\bf 0} & {\bf 1} & M^{T}_{\rm S}M^{-1}_{\mu} \cr {\bf 0} &
\left(M^{T}_{\rm S}M^{-1}_{\mu}\right)^{\dag}_{} & {\bf 1}
\end{matrix} \right) \;,\nonumber\\
\mathcal{W}^{\dag}_{}\mathcal{M}\mathcal{W}^{*}_{} &\approx& \left(
\begin{matrix} {\bf 0} & {\bf 0} & M^{}_{\rm D} M^{-1}_{\rm R}
M^{}_{\rm S} \cr {\bf 0} & M^{}_{\rm R} & {\bf 0} \cr M^{T}_{\rm S}
M^{-1}_{\rm R} M^{T}_{\rm D} & {\bf 0} & {\bf 0}
\end{matrix} \right) \;,
\end{eqnarray}
respectively. One can further derive that the left-handed and
additional gauge-singlet neutrinos have nearly degenerate masses
$M^{}_{\nu} \sim M^{}_{\rm D} M^{-1}_{\rm R} M^{}_{\rm S}$ and form
the pseudo-Dirac particles. However, since $M^{}_{\rm S}\gg
M^{}_{\rm D}$, one has to require that the scale of $M^{}_{\rm R}$
in the GISM be even higher than that in the type-I seesaw model,
which is unnatural.

\begin{figure*}[!ht]
  \begin{center}
  \includegraphics[width=1\textwidth]{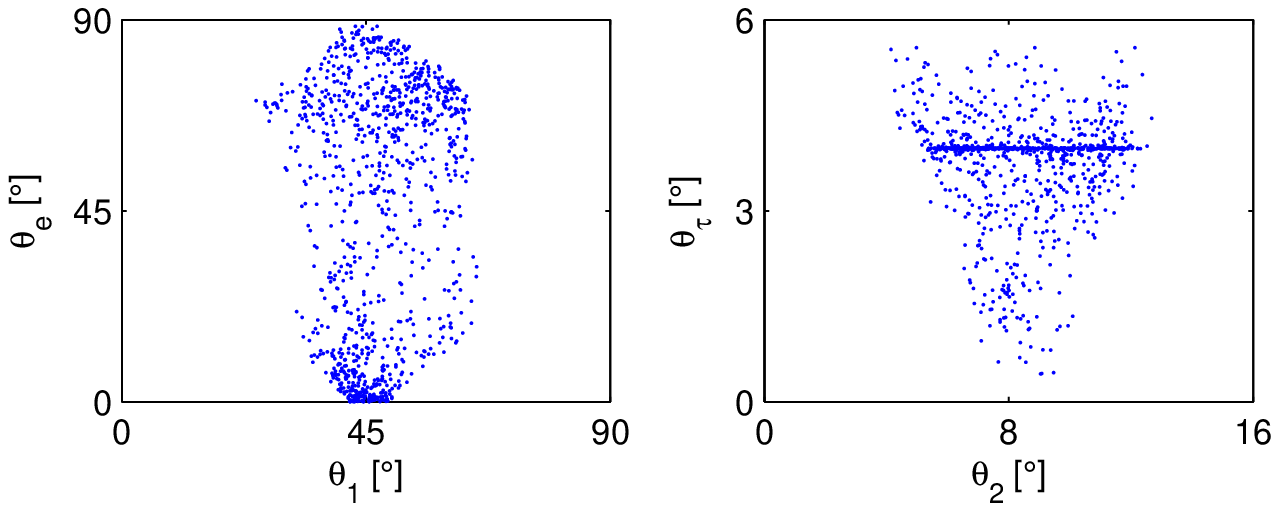} \\
  \end{center}
  \vspace{-.8cm}
  \caption{The comparison of the values between $\theta^{}_1$
  and $\theta^{}_e$ (left) and that of the values between $\theta^{}_2$ and
  $\theta^{}_\tau$ (right). The free parameters $\theta^{}_{e}$,
  $\theta^{}_{3}$, $\delta^{}_{\mu}$, $\delta^{}_{2}$, $\alpha$,
  $\beta$ and $m^{}_1$ are used as inputs.
  The constraints are given by $\Delta m^2_{21} \simeq
(7.4 - 7.8)\times 10^{-5}_{}\; {\rm eV}^2_{}$, $\Delta m^2_{31}
\simeq (2.4 - 2.7) \times 10^{-3}_{}\; {\rm eV}^2_{}$,
$\theta^{}_{23}\simeq (42^\circ_{}-49^\circ_{})$,
$\theta^{}_{12}\simeq (33^\circ_{}-35^\circ_{})$ and
$\theta^{}_{13}\simeq (8.0^\circ_{}-9.6^\circ_{})$. }
\end{figure*}

\begin{figure*}[!ht]
\hspace{-0.2cm}
  \begin{center}
  \includegraphics[width=1\textwidth]{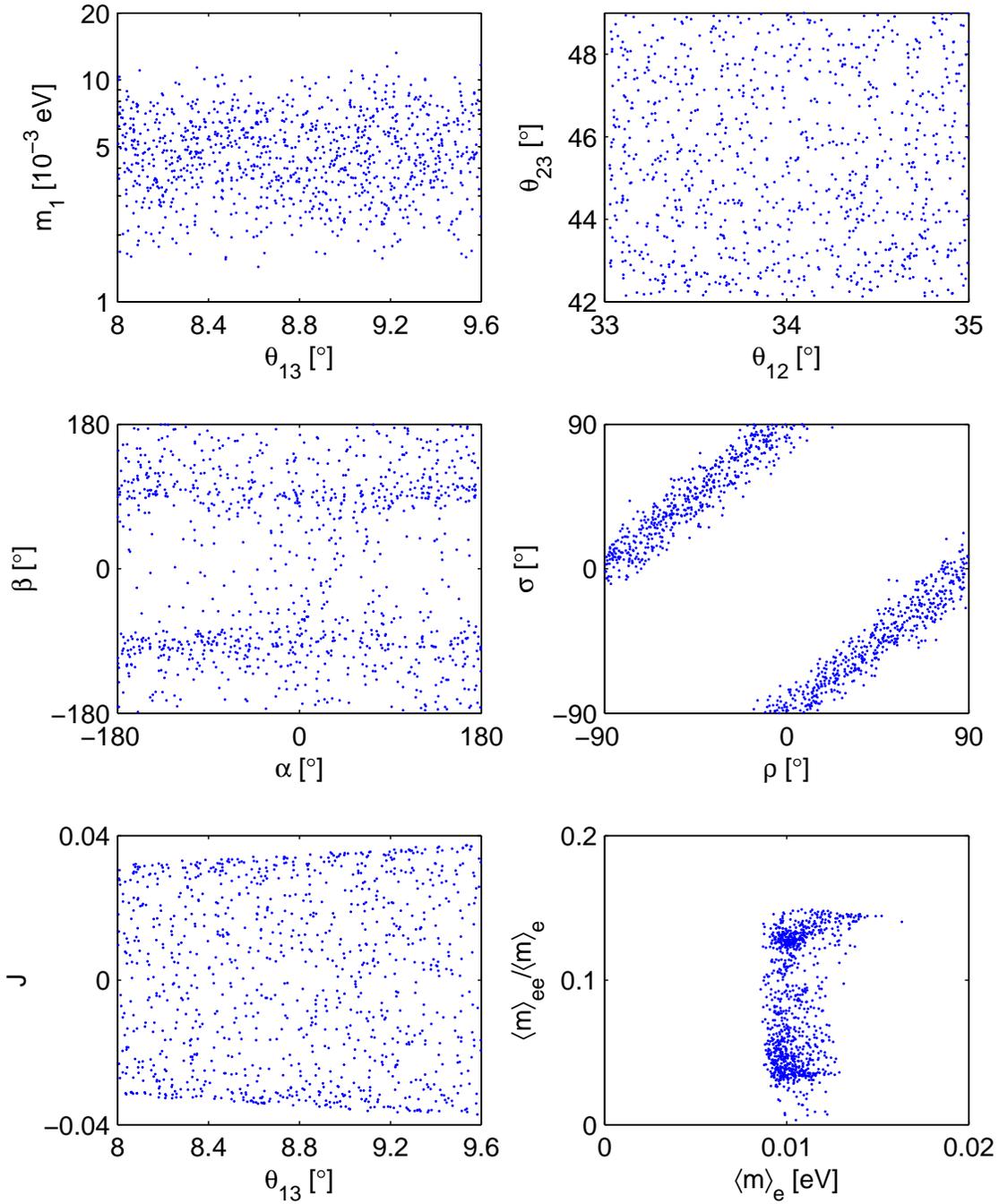} \\
  \end{center}
  \vspace{-.8cm}
  \caption{The parameter space and phenomenological
  predictions in the general case.
  The inputs and constraints are the same as in Fig. 1.  }
\end{figure*}

\begin{figure*}[!ht]
\hspace{-0.2cm}
  \begin{center}
  \includegraphics[width=1\textwidth]{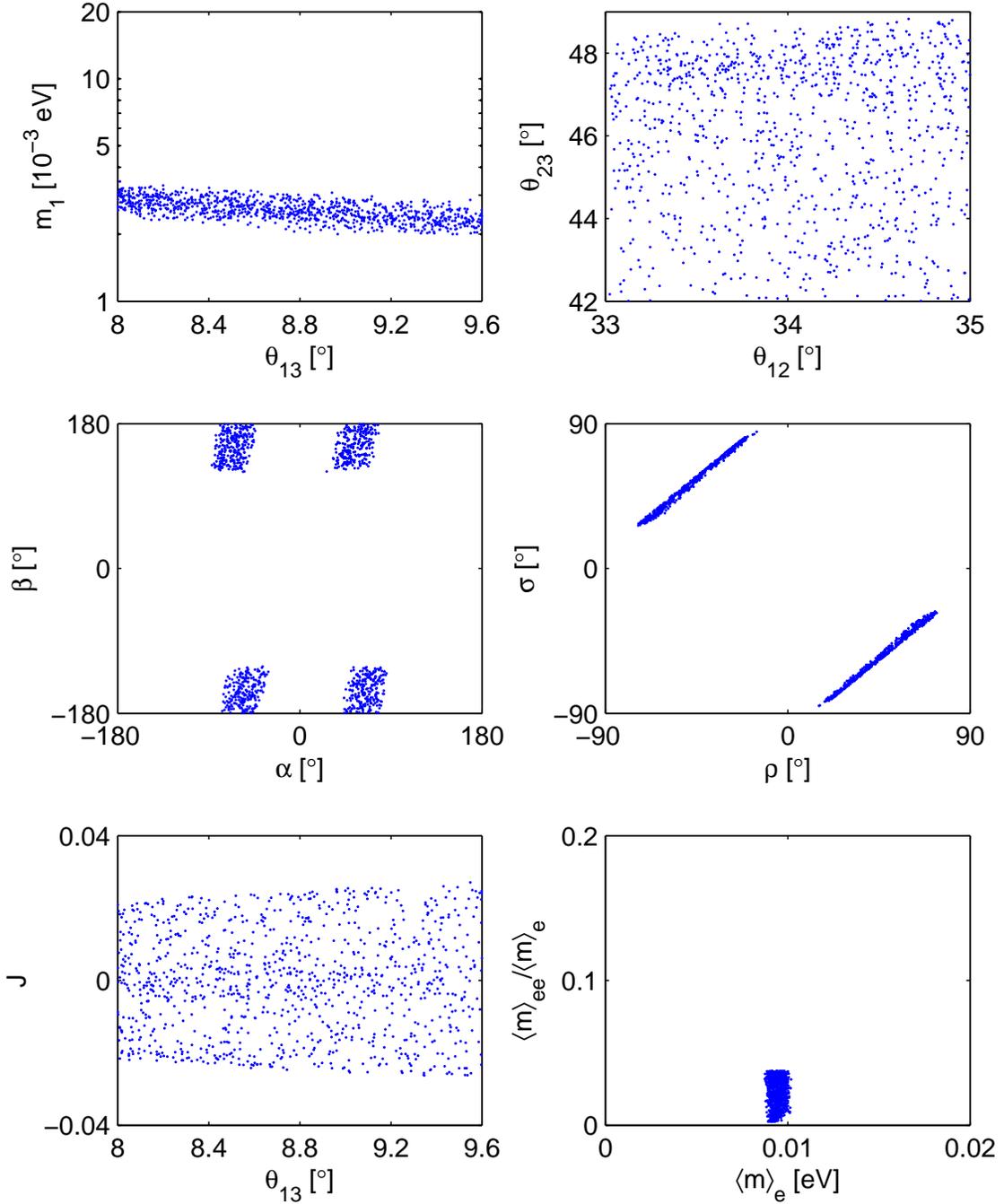} \\
  \end{center}
  \vspace{-.8cm}
  \caption{The parameter space and phenomenological
  predictions of ansatz (A). Only three free parameters $\alpha$,
  $\beta$ and $m^{}_1$ are adjustable.
  The constraints are the same as in Fig. 1.}
\end{figure*}

\begin{figure*}[!ht]
\hspace{-0.2cm}
  \begin{center}
  \includegraphics[width=1\textwidth]{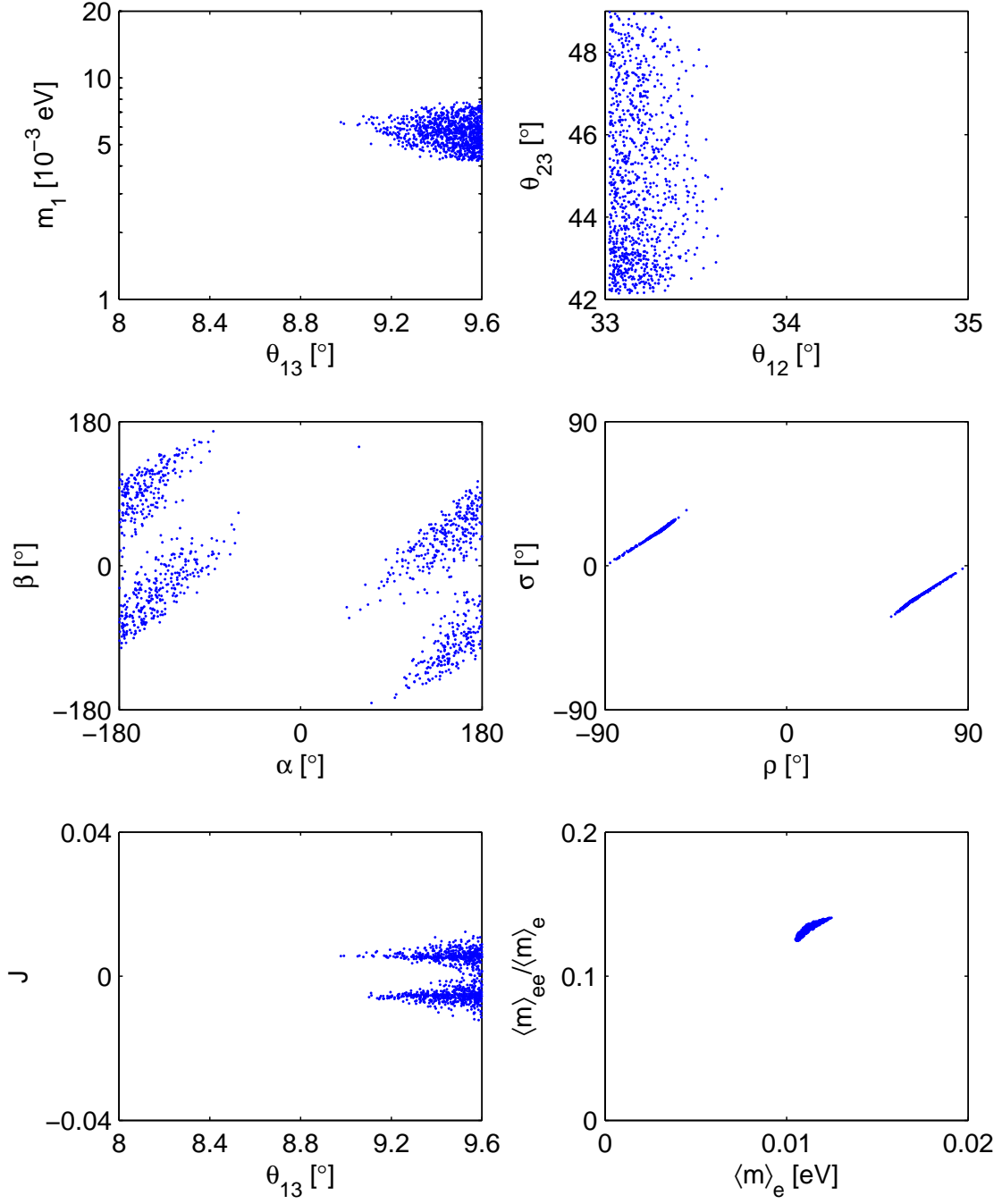} \\
  \end{center}
  \vspace{-.8cm}
  \caption{The parameter space and phenomenological
  predictions of ansatz (B). The inputs and constraints
  are the same as in Fig. 3.}
\end{figure*}

\end{document}